\documentclass[twocolumn,english,aps,prl,twocolumn,superscriptaddress,showpacs,floatfix]{revtex4-1}
\usepackage{mathptmx}

\usepackage[T1]{fontenc}
\usepackage[latin9]{inputenc}
\usepackage{color}
\usepackage{amsmath}
\usepackage{amssymb}
\usepackage{graphicx}

\makeatletter

\providecommand{\tabularnewline}{\\}

\makeatother

\usepackage{babel}
\begin{document}

\title{Kitaev interactions of the spin-orbit coupled magnet UO$_{2}$}

\author{Joseph A. M. Paddison}
\altaffiliation{\href{mailto:paddisonja@ornl.gov}{paddisonja@ornl.gov}} 
\affiliation{Neutron Scattering Division, Oak Ridge National Laboratory, Oak Ridge,
Tennessee 37831, USA}
\author{Lionel Desgranges}
\affiliation{Energy Division, CEA, Cadarache, F 13108, St Paul Lez Durance, France}

\author{Gianguido Baldinozzi}
\affiliation{Centralesup{\'e}lec, Centre National de la Recherche Scientifique, Structures
Property and Modeling of Solids Laboratory, Universit{\'e} Paris-Saclay,
Gif-sur-Yvette 91190, France}

\author{Gerard H. Lander}
\affiliation{Institut Laue-Langevin, 71 avenue des Martyrs, CS 20156, 38042 Grenoble
cedex 9, France}

\author{Henry E. Fischer} 
\altaffiliation{\href{mailto:fischer@ill.fr}{fischer@ill.fr}} \affiliation{Institut Laue-Langevin, 71 avenue des Martyrs, CS 20156, 38042 Grenoble cedex 9, France}

\begin{abstract}
Uranium dioxide, UO$_{2}$, is a canonical example of a magnetic material
with strong spin-orbit coupling. Here, we present a study of the magnetic
diffuse scattering measured on a polycrystalline sample of UO$_{2}$,
which we interpret in terms of its magnetic interactions between U$^{4+}$
magnetic moments. By refining values of the magnetic interaction parameters
to magnetic diffuse-scattering data measured above the magnetic ordering
transition temperature, we show that the dominant magnetic coupling
in UO$_{2}$ is a bond-dependent interaction analogous to the Kitaev
model of honeycomb magnets. We compare our experimental results with
published theoretical predictions and experimental measurements of
the magnetic excitation spectrum. Our results suggest that magnetic
materials with $f$-electron magnetic ions, particularly actinides,
may be promising candidates for realising Kitaev magnetism, and highlight
the role that magnetic diffuse-scattering data can play in identifying
such materials.
\end{abstract}

\maketitle

\begin{figure*}
\begin{centering}
\includegraphics[scale=1.25]{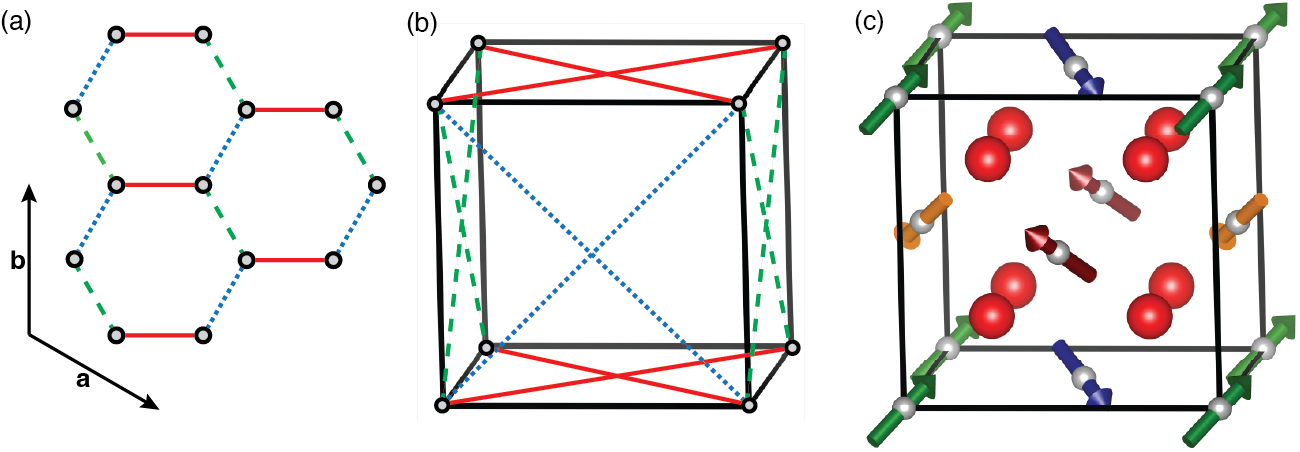}
\par\end{centering}
\centering{}\caption{\label{fig:fig1}(a) Kitaev model on the honeycomb lattice. (b) Kitaev
model on the face-centered cubic lattice. In (a) and (b), the Kitaev
interaction couples the $z$ components of spins connected by red
solid lines, the $x$ components of spins coupled by the green dashed
lines, and the $y$ components of spins coupled by the blue dotted
lines. (c) Transverse triple-$\mathbf{k}$ magnetic structure of UO$_{2}$,
showing U$^{4+}$ ions as grey spheres, O$^{2-}$ ions shown as red
spheres, and spin directions as arrows. Different colours of the spins
are used to show the four $\langle111\rangle$ directions along which
spins are aligned in this non-coplanar magnetic structure.}
\end{figure*}

\affiliation{Institut Laue-Langevin, 71 avenue des Martyrs, CS 20156, 38042 Grenoble
cedex 9, France}
The Heisenberg model of isotropic magnetic interactions is among the
most important models in magnetism \citep{Skomski_2008}. Recently,
however, there has been renewed interest in magnetic interactions
that deviate from the isotropic Heisenberg limit and are strongly
bond-dependent. This interest has been motivated by the theoretical discovery
of a topological spin-liquid ground state for the Kitaev model on
the honeycomb lattice {[}Figure~\ref{fig:fig1}(a){]}, which provides
a foundational example of bond-dependent magnetism \cite{Kitaev_2003,Rousochatzakis_2018}
that has potential applications in quantum computing \citep{Baskaran_2007}.
Subsequently, the Kitaev model has been generalized to other lattices
\citep{Kimchi_2014,Chaloupka_2015}, including triangular, pyrochlore
and face-centred cubic {[}Figure~\ref{fig:fig1}(b){]}. While these
lattices do not show equivalent ground states to the honeycomb Kitaev
model, they can nevertheless exhibit many unusual magnetic properties,
including spin-liquid behaviour \citep{Kato_2017,Kos_2017} and multipolar
excitations \citep{Bai_2021}. Moreover, identifying Kitaev materials
on any lattice can help to understand the factors that enhance the
Kitaev interaction \citep{Winter_2017}. Consequently, identifying
materials where Kitaev interactions are dominant is an important goal
in condensed-matter physics. 

A key ingredient to stabilise large Kitaev interactions is spin-orbit
coupling, which is enhanced for heavier magnetic elements. Motivated
by this observation, we decided to re-examine the magnetic interactions
of uranium dioxide, UO$_{2}$. Uranium dioxide has been extensively
studied due to its use as a nuclear-fuel material at high temperatures
\citep{Hurley_2022,Desgranges_2023}, but it is also of fundamental interest due to
its low-temperature properties \citep{Santini_2009}. At low temperatures,
strong coupling of spin, orbital and lattice degrees of freedom has
been revealed by the observation of mixed magnon-phonon excitations
\citep{Caciuffo_1999,Carretta_2010,Caciuffo_2011}, and by magnetoelastic
responses with potential applications in magnetic sensing \citep{Schonemann_2021,Tereshina-Chitrova_2024}.
Uranium dioxide crystallises in the fluorite structure (space group
$Fm\bar{3}m$; $a=5.46$\,\AA) \citep{Willis_1964}, where U$^{4+}$
ions occupy a face-centred cubic lattice. A first-order transition
to an antiferromagnetic state occurs at $T_{\mathrm{N}}=30.8$\,K
\citep{Willis_1965,Frazer_1965}. The small value of $T_{\mathrm{N}}$
compared to the antiferromagnetic Weiss temperature $\theta\sim-220$\,K
\citep{Nasu_1966} suggests that magnetic interactions are frustrated.
The magnetic propagation vector $\mathbf{k}=[001]$. While neutron-diffraction
experiments were initially interpreted in terms of a single-\textbf{k}
antiferromagnet with the spin direction transverse to $\mathbf{k}$
\citep{Faber_1976}, subsequent measurements established that the
magnetic ground state is actually a transverse triple-\textbf{k} structure
\citep{Blackburn_2005} that can be described by a superposition of
the spin orientations within single-\textbf{k} domains {[}Figure~\ref{fig:fig1}(c){]}.

To explain the properties of UO$_{2}$, a theoretical framework has
been developed that includes crystal-field effects and interactions
between dipolar and quadrupolar degrees of freedom \citep{Carretta_2010,Caciuffo_2011,Bultmark_2009,Pourovskii_2019,Zhou_2022}.
The crystal-field ground state is a triplet that is separated from
the first excited state by approximately 150\,meV \citep{Amoretti_1989,Zhou_2011,Lander_2021,Sundermann_2025},
so that below room temperature, the magnetic moments can be modelled
with an effective spin quantum number $S=1$. This has two important
implications for the magnetic Hamiltonian. First, axial single-ion
anisotropies are not allowed in a cubic environment for any $S$,
and an effective $S<2$ also prohibits ``cubic'' single-ion anisotropy
terms \citep{Dudzinski_1998}. Consequently, no single-ion anisotropy
is expected in UO$_{2}$. Second, an effective $S=1$ carries both
dipolar and quadrupolar degrees of freedom; e.g., the $S_{z}=0$ state
carries a quadrupole moment but no dipole moment \citep{Bai_2021}.
Although excitations of purely quadrupolar degrees of freedom are
silent in neutron-scattering measurements, it proved possible to observe
hybridised dipolar-quadrupolar excitations in UO$_{2}$ \citep{Carretta_2010,Caciuffo_2011},
and these measurements yielded deep insights into the exchange interactions
of both dipolar and quadrupolar origin. Due to the complexity of the
low-temperature excitation spectrum, however, a direct refinement
of the magnetic interactions to experimental data has not yet been
available. 

Here, we report experimentally-derived values of the magnetic exchange
interactions in UO$_{2}$ based on fits to magnetic diffuse-scattering
data measured on a polycrystalline sample above $T_{\mathrm{N}}$.
We find that the magnetic diffuse scattering data cannot be described
by an isotropic Heisenberg model; instead, a good description of our
data requires a large Kitaev interaction. Our fitted values of the
dipolar interactions are in qualitative agreement with the most recent
theoretical predictions \citep{Bultmark_2009,Pourovskii_2019}. We
discuss our results in the context of the low-temperature excitation
spectra \citep{Caciuffo_1999}, and compare UO$_{2}$ with other materials
that realize bond-dependent magnetism on the FCC lattice.

Our UO$_{2}$ sample was comprised of two cylindrical polycrystalline ingots, 
each of dimensions 8.3\,mm in diameter and 14\,mm in height, stacked on
top of each other within a sealed thin-walled vanadium sample container of
diameter 9\,mm.  The two UO$_{2}$ ingots were obtained by sintering at 1970\,K 
under a 5\% Ar/H$_2$ atmosphere, and contained about 0.1\,mol.\% impurities. Neutron diffraction data \citep{D4data_2018}
were measured at temperatures between $10$ and $300$~K 
using the D4 instrument \citep{Fischer_2002} at the Institut Laue-Langevin 
(Grenoble, France) with incident wavelength $\lambda=0.5$\,\AA, providing a maximum wavevector transfer of $Q_{\rm max} = 23.5$~\AA$^{-1}$
that is sufficient for a good-quality Fourier transform to a
pair-distribution function PDF($r$).  Using a short neutron
wavelength ensures that the magnetic scattering is integrated 
over a wide energy range, yielding an accurate estimate of the equal-time
spin correlations; however, it also relaxes the $Q$-resolution, resulting
in relatively broad nuclear Bragg peaks. With these effects in mind,
the data were processed as follows. First, each data set was normalised
in absolute intensity units (barn/sr/U) using the limiting values
of the pair-distribution function. It was necessary to normalise each
data set independently, due to thermal-expansion effects that caused
the quantity of sample exposed to the beam to vary slightly with temperature.
Second, to isolate the magnetic scattering arising from the development
of spin correlations below room temperature, a 300-K data set was
subtracted from every other data set. When performing these subtractions,
we accounted for temperature-dependent changes in lattice parameter
by rescaling the $Q$-axis, and changes in thermal motion by dividing
the data by the Debye-Waller factor and subtracting the thermal diffuse
scattering (TDS), using Rietveld refinements to obtain initial estimates
of lattice parameters and Debye-Waller factors, and fits of the high-$Q$
region to a $Q^{2}$ intensity dependence to estimate the TDS. This
process allowed for a relatively clean subtraction of the nonmagnetic
signals, except in the vicinity of the most intense nuclear Bragg
scattering, where it was necessary to exclude some data points.

\begin{figure*}
\includegraphics{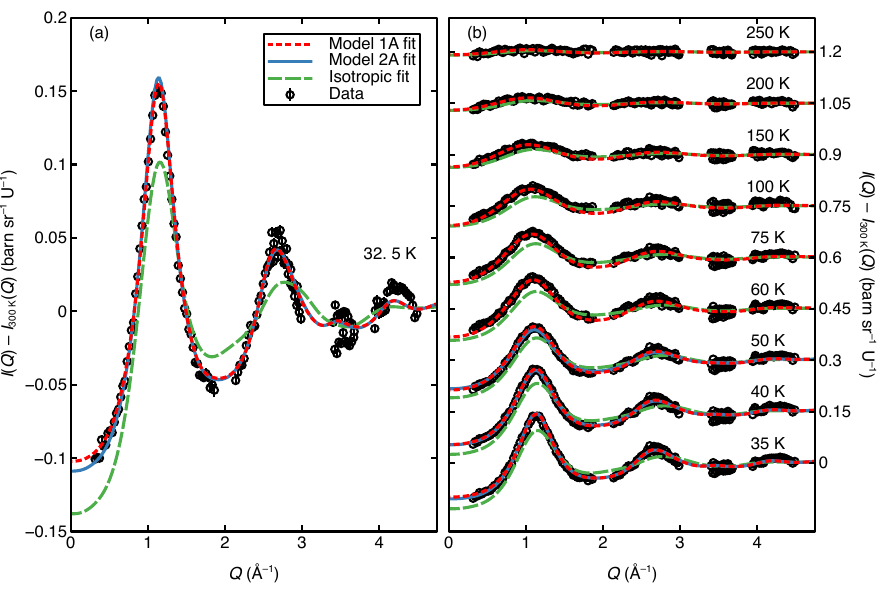}
\centering{}\caption{\label{fig:fig2}Magnetic diffuse scattering data (black circles)
and model fits (lines), showing (a) 32.5\,K data only, and (b) data
measured at 35\,K to 250\,K, with successive temperatures vertically
offset by 0.15 units for clarity. Temperatures are labelled above
each curve and a data set measured at 300\,K has been subtracted
from every data set shown. Green dashed lines show model fits to all
data for isotropic interactions up to fourth neighbours, allowing
an intensity scale factor to vary separately for each data set. Red
dotted lines show model fits to all data for the $J$-$K$-$\Gamma$-$J_{3}$
model (Eq.~(\ref{eq:JK_model})), allowing an intensity scale factor
to vary separately for each data set (Model 1A in Table~\ref{tab:param_values}).
Blue solid lines show model fits to data measured between 32.5\,K
and 50\,K for the $J$-$K$-$\Gamma$-$J_{3}$ model (Eq.~(\ref{eq:JK_model})),
allowing a common intensity scale factor for all data sets (Model
2A in Table~\ref{tab:param_values}). Fits for Model 2A and Model
2B resemble fits for Model 1A and Model 1B, respectively, and are
not shown.}
\end{figure*}

The magnetic diffuse scattering data measured above $T_{\mathrm{N}}$
are shown in Figure~\ref{fig:fig2}. They show the development of
a peak centred around $1$\,\AA$^{-1}$ as the sample is cooled
from room temperature to just above $T_{\mathrm{N}}$, indicating
the development of antiferromagnetic correlations with an increasing
coherence length. These data were previously published \citep{Desgranges_2025}
and analysed using the reverse Monte Carlo method \citep{Paddison_2013},
revealing antiferromagnetic correlations with magnitudes consistent
with a theoretical study \citep{Pourovskii_2019}. The magnetic diffuse
scattering can be directly related to the exchange interactions by
applying a reaction-field approximation \citep{Brout_1967,Logan_1995,Paddison_2020,Paddison_2023}.
For a structure such as UO$_{2}$, which has a single magnetic 
atom in its primitive unit cell, the Fourier transform of the magnetic
interactions is written as a $3\times3$ matrix $\mathsf{J}(\text{\textbf{Q}})$
with elements $J_{\alpha\beta}(\mathbf{Q})=-\sum_{\mathbf{R}}J_{\alpha\beta}(\mathbf{R})e^{-i\mathbf{Q}\cdot\mathbf{R}}$,
where $\alpha,\beta$ are Cartesian spin components, and $J_{\alpha\beta}(\mathbf{R})$
is the coefficient of $S_{i}^{\alpha}S_{j}^{\beta}$ in Eq.~(\ref{eq:heisenberg}),
for pair of spins $i,j$ that are separated by a primitive lattice
vector $\mathbf{R}$. The interaction matrix is diagonalised to obtain
its eigenvalues $\lambda_{\mu}(\mathbf{Q})$ and eigenvector components
$U_{\mu}^{\alpha}(\mathbf{Q})$. The magnetic diffuse scattering intensity
is obtained as
\begin{equation}
I(\mathbf{Q})=C\frac{[\mu_{\mathrm{eff}}f(Q)]^{2}}{3}\sum_{\mu=1}^{3}\frac{|\mathbf{s}_{\mu}^{\perp}(\mathbf{Q})|^{2}}{1-\chi_{0}\left[\lambda_{\mu}(\mathbf{Q})-\lambda\right]},\label{eq:intensity_mfo}
\end{equation}
where $\mu_{\mathrm{eff}}$ is the magnetic moment per ion, $C=0.07265$~barn,
$f(Q)$ is the U$^{4+}$ magnetic form factor given in Ref.~\citep{Lander_1976},
$\chi_{0}=1/3T$ is the Curie susceptibility, and for the FCC lattice
$\mathbf{s}_{\mu}^{\perp}(\mathbf{Q})=\sum_{\alpha=1}^{3}(\hat{\mathbf{n}}_{\alpha}-\mathbf{Q}\thinspace\hat{\mathbf{n}}_{\alpha}\cdot\mathbf{Q}/Q^{2})U_{\mu}^{\alpha}(\mathbf{Q})$,
where $\hat{\mathbf{n}}_{\alpha}$ is a unit vector parallel to $\mathbf{x},\mathbf{y},$
or $\mathbf{z}$ \citep{Paddison_2020,Paddison_2023}. Magnetic diffuse
scattering calculations were performed using the
Spinteract program \citep{Paddison_2023}, which numerically powder
averages Eq.~(\ref{eq:intensity_mfo}) and convolves the result with
the instrumental resolution function of D4. This procedure allows
for a quantitative comparison of the calculated diffuse scattering
patterns with our experimental data.

As a first attempt to analyse these data in terms of the magnetic
Hamiltonian, we considered the isotropic Heisenberg model,
\begin{align}
H_{\mathrm{iso}} & =\sum_{\left\langle i,j\right\rangle \in n}J_{n}\mathbf{S}_{i}\cdot\mathbf{S}_{j},\label{eq:heisenberg}
\end{align}
where $J_{n}$ is the isotropic exchange interaction between $n$-th
neighbours, and we employ the conventions that antiferromagnetic interactions
have positive sign, each pairwise interaction is counted once, and
the spins are classical vectors of length $\sqrt{S(S+1)}=\sqrt{2}$.
We attempted to refine the values of the four nearest-neighbour couplings
against the data collected between 31 and 250 K simultaneously. Figure~\ref{fig:fig2}
shows a comparison of our experimenal data with the best fit for this
model. The isotropic model does not describe the
data well, since it fails to capture the shape of the main diffuse
peak at temperatures below 150 K. In the refinements shown, an overall
intensity scale factor was allowed to refine independently for each
data set; hence, the disagreement cannot be addressed by a simple
scaling of the fitted curves. Attempts to refine additional parameters---including
magnetic interaction parameters up to $n=7$---also failed to produce
a meaningful improvement in the fit quality. We conclude from this
that our data cannot be explained by an isotropic model.

We therefore consider a more general Hamiltonian that contains all
nearest-neighbour interactions allowed by the space-group symmetry
of UO$_{2}$, but is restricted to only bilinear interactions; i.e.,
those that are linear in components of each spin. This Hamiltonian
can be written as
\begin{align}
H_{\mathrm{bd}} & =\sum_{\left\langle i,j\right\rangle \in n}J_{n}\mathbf{S}_{i}\cdot\mathbf{S}_{j}+K\sum_{\left\langle i,j\right\rangle _{\gamma}}S_{i}^{\gamma}S_{j}^{\gamma}+\Gamma\sum_{\left\langle i,j\right\rangle _{\gamma}}\left(S_{i}^{\alpha}S_{j}^{\beta}+S_{i}^{\beta}S_{j}^{\alpha}\right),\label{eq:JK_model}
\end{align}
where the bond-dependent interactions are $K$ (the Kitaev interaction)
and $\Gamma$ \citep{Cook_2015,Diop_2022,Paddison_2024}. The notation
$\left\langle i,j\right\rangle _{\gamma}$ indicates that the sum
runs over pairs of U$^{4+}$ ions in the plane perpendicular to the $\gamma$-axis;
for example, if we consider a pair of U$^{4+}$ ions within the $xy$ plane, then the Kitaev term couples their $z$ components
$S_{i}^{z}S_{j}^{z}$, whereas the $\Gamma$ term couples their $x$
and $y$ spin components, $S_{i}^{x}S_{j}^{y}+S_{i}^{y}S_{j}^{x}$.
The corresponding $\mathsf{J}(\text{\textbf{Q}})$ is given explicitly
in Ref.~\citep{Cook_2015}.

\begin{table*}
\begin{centering}
\begin{tabular}{c|ccccc|ccc}
\hline 
 & $J$ (K) & $K$ (K) & $\Gamma$ (K) & $J_{2}$ (K) & $J_{3}$ (K) & $R_{\mathrm{wp}}$ (\%) & $T_{\mathrm{N}}^{\mathrm{calc}}$ (K) & $\mu_{\mathrm{eff}}$ ($\mu_{\mathrm{B}}$)\tabularnewline
\hline 
Model 1A (independent scale, $T\leq250$\,K) & $2.4(2)$ & $19.3(3)$ & $-9.0(3)$ & $0^{*}$ & $0.13(2)$ & $19.8$ & $19.5$ & 2.01 - 2.88\tabularnewline
Model 1B (independent scale, $T\leq250$\,K) & $2.5(2)$ & $17.6(3)$ & $-8.4(2)$ & $-0.6(1)$ & $0^{*}$ & $19.7$ & $20.0$ & 2.03 - 2.96\tabularnewline
Model 2A (common scale, $T\leq50$\,K) & $5.6(2)$ & $14.5(3)$ & $-7.1(2)$ & $0^{*}$ & 0.15(2) & $15.4$ & $19.3$ & 2.08\tabularnewline
Model 2B (common scale, $T\leq50$\,K) & $5.3(2)$ & $13.8(3)$ & $-6.6(2)$ & $-0.3(1)$ & $0^{*}$ & $15.5$ & $16.2$ & 2.11\tabularnewline
\hline 
Theory (Ref.~\citep{Pourovskii_2019}) & $8.2$ & $14.1$ & $-3.9$ & $+0.8$ & $0^{*}$ &  &  & \tabularnewline
Theory (Ref.~\citep{Bultmark_2009}) & $3.0$ & $6.9$ & $0^{*}$ & $0^{*}$ & $0^{*}$ &  &  & \tabularnewline
Experiment (Ref.~\citep{Caciuffo_2011}) & $4.5$ & $13.5$ & $0^{*}$ & $0^{*}$ & $0^{*}$ &  &  & \tabularnewline
\hline 
\end{tabular}
\par\end{centering}
\caption{\label{tab:param_values}Refined interaction parameter values, sum
of squared residuals $R_{\mathrm{wp}}$, calculated magnetic ordering
temperature $T_{\mathrm{N}}$, and calculated effective paramagnetic
moment $\mu_{\mathrm{eff}}$. Literature results are shown for comparison,
which are reproduced from Ref.~\citep{Pourovskii_2019} and converted
to match the convention of Eq.~(\ref{eq:JK_model}). Values denoted
with an asterisk ({*}) were fixed during the refinements.}
\end{table*}

Theoretical studies have shown that bond-dependent couplings are important
in determining the magnetic ground state on the FCC lattice \citep{Cook_2015,Diop_2022}.
For UO$_{2}$, Heisenberg interactions are antiferromagnetic ($J>0$);
this supports antiferromagnetic ordering with $\mathbf{k}=[100]$,
as is observed experimentally. In this regime, the sign of the Kitaev
term plays a central role in determining the specific ground state
\citep{Cook_2015,Diop_2022}. Positive values of $K$ stabilise a
ground state with the spin direction perpendicular to $\mathbf{k}$,
whereas negative values of $K$ stabilise a ground state with spins
parallel to $\mathbf{k}$. The effect of $\Gamma$ on the ground state
is more subtle, but $\mathbf{k}=[100]$ ordering remains stable for
$K<0$ and relatively small $|\Gamma|\lesssim K$. An unusual feature
of the Hamiltonian defined by Eq.~(\ref{eq:JK_model}) is that single-$\mathbf{k}$
and triple-$\mathbf{k}$ structures are energetically degenerate for
any choice of $J$, $K$ and $\Gamma$ that stabilises magnetic ordering
with $\mathbf{k}=[100]$ \citep{Diop_2022,Paddison_2024}. This implies
that further interactions are required to stabilise the triple-$\mathbf{k}$
ordering observed in UO$_{2}$. We will return to this point below.

Figure~\ref{fig:fig2} shows fits to our neutron data allowing four
interaction parameters to vary---$J$, $K$, $\Gamma$, and an isotropic
third-neighbour coupling $J_{3}$. Excellent agreement is now obtained
for this model---which contains the same number of free parameters
as the isotropic model considered above---suggesting that anisotropic
interactions are important in UO$_{2}$. Initially, we allowed an
overall intensity scale factor to refine independently for each data
set; the values obtained indicated a significant temperature dependence
of the effective magnetic moment, from approximately 2.0$\,\mu_{\mathrm{B}}$
at 31~K to 2.9$\,\mu_{\mathrm{B}}$ at 250~K, which may be related
to thermal population of excited crystal-field levels. A reasonable
refinement could be achieved with a common scale factor for data collected
between 31 and 50\,K, so a second set of refinements was performed
over this restricted temperature range. Refinements were also performed
allowing the isotropic next-nearest-neighbour coupling $J_{2}$ to
vary instead of $J_{3}$. For each model, Table~\ref{tab:param_values}
presents the fitted parameter values, the sum of squared residuals
$R_{\mathrm{wp}}$, and the calculated magnetic ordering temperature
$T_{\mathrm{N}}^{\mathrm{calc}}$.

We now discuss the physical implications of the results given in Table~\ref{tab:param_values}.
While different refinement protocols yielded significant differences
in the refinement parameter values, several conclusions can be reached.
Most importantly, the Kitaev interaction is always positive and is
the largest term, with a magnitude at least double that of any other
interaction. A key result of our work is therefore that $K$ provides
the dominant energy scale in UO$_{2}$. The positive sign of $K$
is consistent with the observed transverse magnetic structure \citep{Diop_2022}.
The next-largest interaction is $\Gamma$, which our refinements show
to have a negative value. The Heisenberg interaction is antiferromagnetic,
as anticipated, but significantly smaller than $K$. The further neighbour
couplings are the smallest terms, and our data are nearly equally
well described by a small ferromagnetic $J_{2}$ (models 1B, 2B) or antiferromagnetic
$J_{3}$ (models 1A, 2A). Either of these possibilities can help to stabilise $\mathbf{k}=[100]$
ordering, as observed experimentally, whereas the opposite choice
of signs (ferromagnetic $J_{3}$ or antiferromagnetic $J_{2}$) would
stabilise ordering with $\mathbf{k}=[1\frac{1}{2}0]$ \citep{Diop_2022}.
The calculated magnetic ordering temperatures are typically around
20\,K, which is a significant ($\sim$30\%) underestimate compared
to the experimental value of 30.8~K. This may be due to the use of
the reaction-field approximation, which can underestimate magnetic
ordering temperatures by up to $\sim$20\% \citep{Wysin_2000}. It
may also occur because the magnetic transition in the real material
is first-order and driven by quadrupolar interactions, which are not
considered in Eq.~(\ref{eq:JK_model}). Overall, the results shown
in Table \ref{tab:param_values} suggest the following range of values
for the interaction parameters: $2\lesssim J\lesssim6\,\textrm{K}$,
$14\lesssim K\lesssim20\,\textrm{K}$, $-9\lesssim\Gamma\lesssim-6\,\textrm{K}$,
$-0.7\lesssim J_{2}\lesssim0\,\textrm{K}$, and $0\lesssim J_{3}\lesssim0.2\,\textrm{K}$.

Estimates of all the magnetic interactions in Eq.~(\ref{eq:JK_model})
have been reported from theory, and some terms have also been previously
estimated experimentally. Our results are consistent with the values
of $J$ and $K$ estimated in a neutron spectroscopy study, in which
$\Gamma$ was not reported \citep{Caciuffo_2011}. A detailed comparison
with early inelastic scattering work \citep{Cowley_1968} is not possible
due to its assumption of a single-$\mathbf{k}$ structure, but the
overall energy scales we identify are qualitatively consistent. Our
results are also in remarkably good agreement with a recent first-principles
study \citep{Pourovskii_2019}, showing the same signs and similar
magnitudes of $J$, $K$, and $\Gamma$. Agreement in the relative
magnitudes of $J$ and $K$ is also obtained with an independent theoretical
study \citep{Bultmark_2009}. Our experimental analysis therefore
provides strong support for the theoretical results of Refs.~\citep{Pourovskii_2019,Bultmark_2009}. 

\begin{figure*}
\begin{centering}
\includegraphics{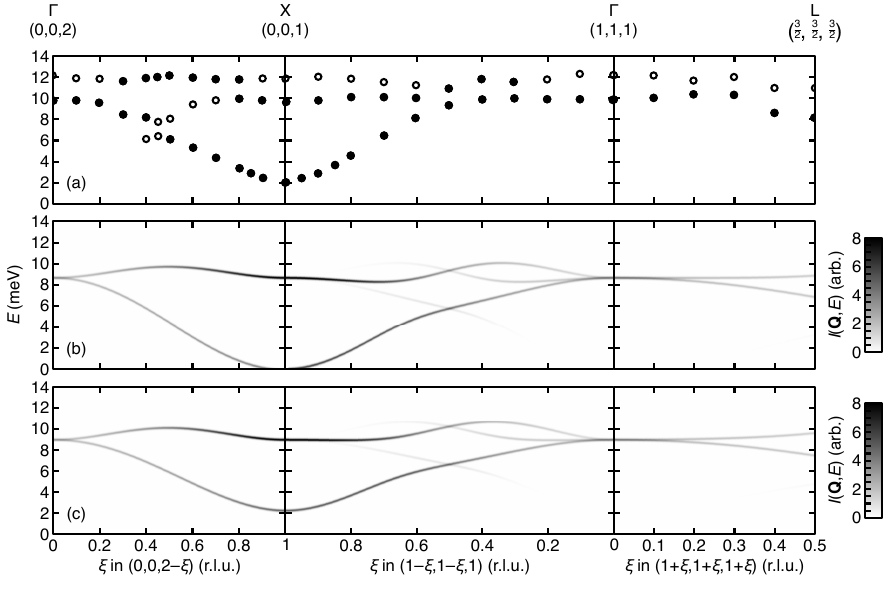}
\par\end{centering}
\centering{}\caption{\label{fig:fig3}Magnetic excitations in UO$_2$, where r.l.u. denotes reciprocal-lattice units. (a) Previously published dispersion relation for
UO$_{2}$, reproduced from Ref.~\citep{Caciuffo_1999}. (b) Calculated
inelastic neutron-scattering intensity for the $J$-$K$-$\Gamma$-$J_{3}$
model (Model 1A in Table \ref{tab:param_values}). (c) Calculated
inelastic neutron-scattering intensity for the same model as (b),
except a biquadratic interaction $J_{\mathrm{bq}}=10$\,K is assumed.}
\end{figure*}

Magnetic diffuse-scattering data are primarily sensitive to bilinear
interactions \citep{Bai_2021}. This has the advantage of simplifying
the estimation of the bilinear terms; however, it is important to
investigate how the neglect of quadrupolar interactions affects the
values of the bilinear terms obtained in this way. To do so, we note
that quadrupolar interactions are biquadratic when expressed in terms
of spin components, and consider the simplest case of isotropic nearest-neighbour
biquadratic interactions,
\begin{equation}
H_{\mathrm{bq}}=J_{\mathrm{bq}}\sum_{\left\langle i,j\right\rangle }(\mathbf{S}_{i}\cdot\mathbf{S}_{j})^{2},\label{eq:biquadratic}
\end{equation}
where $J_{\mathrm{bq}}$ denotes the biquadratic exchange coupling.
This is the lowest-order interaction that can remove the degeneracy
between single-$\mathbf{k}$ and triple-$\mathbf{k}$ structures:
positive values of $J_{\mathrm{bq}}$ stabilise triple-$\mathbf{k}$
ordering, whereas negative values of $J_{\mathrm{bq}}$ stabilise
single-$\mathbf{k}$ ordering \citep{Paddison_2024}. In a mean-field
approximation, $J_{\mathrm{bq}}$ has an equivalent effect on the
magnetic diffuse scattering as an isotropic Heisenberg exchange term
\citep{Lines_1966}; hence, a biquadratic coupling shifts fitted values
of $J$. Specifically, the values of $J$ given in Table~\ref{tab:param_values}
can be interpreted as $J=J_{\mathrm{true}}-\frac{1}{2}J_{\mathrm{\mathrm{bq}}}$,
where $J_{\mathrm{true}}$ is the true isotropic bilinear coupling
\citep{Lines_1966}.

As a check on our results, we calculated the magnetic excitation spectrum
for our model parameters using the SpinW program \citep{Toth_2015},
and compared the result with published inelastic-neutron-scattering
measurements of Ref.~\citep{Caciuffo_1999}. Figure~\ref{fig:fig3}(a)
shows the published dispersion relation~\citep{Caciuffo_1999},
which shows a minimum at the $\mathrm{X}=(001)$ point with a gap
of $\approx2$\,meV, and a large gap of $\approx10$\,meV at the
$\Gamma=(002)$ point. Figure~\ref{fig:fig3}(b) shows the calculated
spin-wave spectrum for the $J$-$K$-$\Gamma$-$J_{3}$ model (Model
1A in Table~\ref{tab:param_values}), assuming $S=1$ and the transverse
triple-$\mathbf{k}$ magnetic ground state. The calculation captures
the overall shape of the dispersion curve, including a minimum at
the X point and a large gap at the $\Gamma$ point. This result points
to a large Kitaev term, since for $K=0$, there would be no gap at
the $\Gamma$ point, in strong disagreement with the data. The model also
predicts the observed approximately quadratic spin-wave dispersion around the X point, which
is uncommon for an antiferromagnet. However,
the calculation deviates from the data in several respects: it shows
a gapless spectrum in contrast to the observed $\approx2$\,meV gap
at the X point, the overall bandwidth is too small, and some weaker
modes and splittings at high energy are not reproduced. 
The first of these discrepancies can be addressed by including a nonzero value of $J_{\mathrm{bq}}$.
Figure~\ref{fig:fig3}(c) shows the calculated spin-wave spectrum
assuming $J_{\mathrm{bq}}=10$\,K, which now reproduces the observed
gap. We note that semiclassical spin-wave calculations require the
use of renormalised or shifted values of $J_{\mathrm{bq}}$ and $J$
to obtain correct results \citep{Dahlbom_2022}, which have been included
in our calculation. Our results therefore suggest a substantial biquadratic
interaction of the same order of magnitude as $K$. The $\sim$2\,meV
discrepancy in overall bandwidth and omission of weak high-energy
modes may be related to the omission of anisotropic quadrupolar interactions
and magnon-phonon couplings \citep{Caciuffo_2011,Paolasini_2021},
which are beyond the scope of our work focusing on the bilinear exchange
interactions.

Our study highlights the surprising observation that magnetic diffuse-scattering
data measured on polycrystalline samples are sensitive to bond-dependent
interactions such as the Kitaev coupling \citep{Paddison_2020}. In
UO$_{2}$, our results indicate a dominant Kitaev interaction and
smaller but still substantial $\Gamma$ interaction. The presence of frustrated couplings supports a large temperature range where the system is a correlated paramagnet, where structured magnetic excitations may affect the thermal conductivity \cite{Zheng_2019}. The anisotropic
terms in UO$_{2}$ are much larger than those in FCC magnets where
the magnetic ion is a transition metal. For example, the cubic double
perovskites Ba$_{2}$YRuO$_{6}$ and Ba$_{2}$LuRuO$_{6}$ with magnetic
Ru$^{5+}$ ions (4$d^{3}$) were recently shown to have the same transverse
triple-\textbf{k} ground state as UO$_{2}$, but the ratio $K/J\sim0.02$
is two orders of magnitude smaller in these double perovskites than
in UO$_{2}$ \citep{Paddison_2024}. In K$_{2}$IrCl$_{6}$, in which Ir$^{4+}$ moments (5$d$$^{5}$) also occupy a FCC lattice
and $\mathbf{k}=[1,\frac{1}{2},0]$ is stabilised by fluctuations,
$K/J\sim0.2$ is still an order of magnitude smaller than in UO$_{2}$
\citep{Wang_2025}; a similar ratio has been calculated for Ba$_{\text{2}}$CeIrO$_{6}$
\citep{Revelli_2019}. By contrast, the $4f$-electron magnets CeAs
and CeSb have $|K/J|\sim1.6$, and USb has a very large ratio $|K/J|\sim10$
\citep{Halg_1986}. Taken together with our results for UO$_{2}$,
these observations suggest that materials with magnetic $f$-electron
elements may be the most promising systems to obtain a dominant Kitaev
interaction. This goal remains relevant for lattices such as honeycomb,
where a topologically-nontrivial spin-liquid ground state can only
be realized if the Kitaev term dominates. Candidate materials such
as $\alpha$-RuCl$_{3}$ have been extensively investigated with the
goal in mind, but recent results suggest that the non-Kitaev interactions
are probably substantial \citep{Maksimov_2020}; hence, the search
for ideal Kitaev honeycomb materials continues.

\acknowledgements{Work of J.A.M.P. (data analysis and manuscript writing) was supported by the U.S. Department of Energy, Office
of Science, Basic Energy Sciences, Materials Sciences and Engineering
Division.}


\end{document}